\documentclass[conference,a4paper]{IEEEtran}
\IEEEoverridecommandlockouts
\usepackage{cite}
\usepackage{amsmath,amssymb,amsfonts}
\usepackage{algorithm}
\usepackage{algpseudocode}
\usepackage{graphicx}
\usepackage{textcomp}
\usepackage{blindtext}
\usepackage{subfigure}
\usepackage{float} 
\usepackage{subfigure}
\usepackage{xcolor}
\usepackage{diagbox}
\usepackage{slashbox}
\usepackage{wrapfig}
\usepackage{makecell}
\def\BibTeX{{\rm B\kern-.05em{\sc i\kern-.025em b}\kern-.08em T\kern-.1667em\lower.7ex\hbox{E}\kern-.125emX}}

\makeatletter
\begin{document}
	\title{The Effect of Variable Factors on the Handover Performance for Ultra Dense Network}
	\IEEEpeerreviewmaketitle
	\author{\IEEEauthorblockN{Donglin Wang\IEEEauthorrefmark{2}, Anjie Qiu\IEEEauthorrefmark{2}, Qiuheng Zhou\IEEEauthorrefmark{1}, Sanket Partani\IEEEauthorrefmark{2} and Hans D. Schotten\IEEEauthorrefmark{2}\IEEEauthorrefmark{1}}
		\IEEEauthorblockA{\textit{\IEEEauthorrefmark{2}University of Kaiserslautern, Kaiserslautern, Germany} \\
		$\{$dwang,qiu,partani,schotten$\}$@eit.uni-kl.de \\}
		\IEEEauthorblockA{\textit{\IEEEauthorrefmark{1}German Research Center for Artificial Intelligence (DFKI GmbH), Kaiserslautern, Germany} \\
		$\{$qiuheng.zhou,schotten$\}$@dfki.de}
	}
	\maketitle
	
\begin{abstract}
With wireless communication technology development, the 5G New Radio (NR) has been proposed and developed for a decade. This advanced mobile communication technology has more advancements, such as higher system capacity, higher spectrum efficiency, higher data rates, and so on. In 5G, Ultra-Dense Network (UDN) is deployed for increasing the system capacity and frequency reuse to meet high application requirements. The architecture of 5G UDN is to realize the dense and flexible deployment of smaller general Node B (gNB). However, the increased capacity of applying UDN in 5G is anticipated at the cost of increased signal interference, increased handover times, and increased handover failures. The Time to Trigger (TTT) is one of the most important factors in handover frequency which is deserved to be detected. Moreover, the density of the 5G gNBs influences the handover times and performance as well. In this work, we provide a compendium of 5G handover management. A downlink system-level simulator for 5G handover is built and utilized to evaluate the effect of different TTT values and densities of gNBs on the 5G handover. In addition, different velocities of Traffic Users (TUs) have been applied to the simulation system. From the simulation results, the handover performance has been analyzed and optimized by applying adjustable TTT under different densities of gNBs which will help people have a better understanding of the selection and effect of proper TTT, UDN, and different velocities of TUs on 5G handover performance.
\end{abstract}


\section{introduction}
Nowadays, there is a large growth in mobile data all over the world, especially in connected vehicle areas. In the 3rd Generation Partnership Project (3GPP) releases 15 and 16 [1][2], the 5G NR technology was released with more enhancements. For example, one of the core characteristics of 5G is the UDN to meet the high data traffic requirement. In [3], the IMT-2020 group has identified the UDN as one component of 5G core technologies and as one of the most helpful approaches. By increasing the coverage of 5G gNB, the spectrum efficiency and system capacity are remarkably improved, meanwhile, the number of TUs served by each 5G gNB is dropped which improves the frequency reuse [3]. However, it's clear to see that UDN deployment in 5G will enlarge the handover times. As a consequence, in 5G the capacity is increased but causing damage to increased handover rates and higher signal overheads [4]. 

Before we get to know how the density of 5G gNBs affects the 5G handover rate, we need to know the LTE handover and 5G NR handover. In [5] from LTE to 5G NR, there is a detailed documentary survey on handover management in LTE and 5G NR where many key points regarding handover procedure challenges and techniques are discussed and are required to be considered in formulating an optimized handover scheme. In [6], the authors are providing a comprehensive formal analysis of 5G handover including protocol testing and verification, mobile networking, and so forth. So far, research activities have focused on optimizing the LTE handover scheme [7][8].

TTT is a valuable parameter in the performance of the 5G handover rate. The handover is processed between two neighboring cells if the criterion of handover is met during TTT [9]. The utilization of the TTT time period can prevent excessive frequent handover events in a short time period. But a too large TTT time period may cause connection loss or bad connection quality of TU from one cell to another cell. It's necessary to detect the effect of various TTT values on the handover performance. In [10], Juwon presents a handover optimization scheme for different speeds of TUs in LTE. In this work, adjustable TTT parameters are applied. In this work [11], the effects of TTT values on the handover performance in an LTE system are analyzed. The results from [11] show the handover performance is improved by applying adaptive TTT parameters. However, these works evaluated the effect of various TTT values on the handover performance for LTE handover not for 5G NR handover performance. In our work, we want to evaluate the effect of various TTT values on the 5G handover performance by applying varying densities of gNB from the 5G network. 

The paper is organized as follows: for section II, we are going to show what the 5G handover structure looks like, and how to make a handover-triggering decision. In section III, we establish the 5G UDN simulation scenario and set simulation parameters. Section IV analyzes the simulation results for different simulation scenarios. Lastly, in section V, the conclusion and future work plan are drawn. 

\section{5g handover process}
Generally, in LTE or 5G NR cellular networks, the mobility mechanisms entitle the TU to move within the communication range of networks and be served by networks. There are basically two different states for Radio Access Mobility (RAM).
One is when the TU is in the IDLE\_MODE where the cell selection for the TU happens. The UE in IDLE\_MODE indicates that there is no active data transmission or reception, a fitting cell is required to be selected for incoming data. Subsequently, the UE in IDLE\_MODE will start transmitting and change the state to the active mode or CONNECTED\_MODE. In this work, only TUs in CONNECTED\_MODE are considered. Because handover only happens when the TU is in CONNECTED\_MODE and a new cell is considered better than the current serving cell [5]. 
Research on handover have done so many years for different networks like
[12] for 1G, [13] for 2G, [14] for 3G and [15] for LTE.

In this work, 5G NR handover procedure is considered and is also developed based on LTE handover technologies. 
To be more specific, handovers in 5G networks can be separated into intra-layer-handover and inter-layer-handover. They can be differentiated depending on the serving and targeting networks and whether they are using the same Radio Access Network (RAN) technologies. If the same RAN technology is used for serving and targeting networks, the intra-layer-handover is performed. But if different RAN technologies are used for serving and targeting networks, the inter-layer-handover is needed, such as from 5G to LTE or from LTE to 5G. In this work, we only consider intra-layer-handover from 5G to 5G [16]. 

\begin{figure}[htbp]
	\centering
	\includegraphics[width=\linewidth]{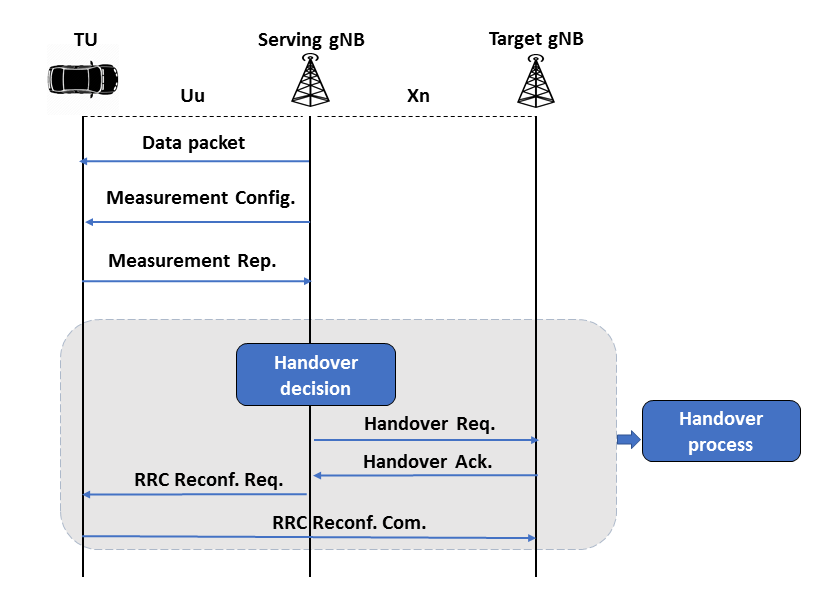}
	\caption{Simplified 5G handover procedure}
	\label{fig}
\end{figure} 

\subsection{Overview of 5G Handover Procedure}
In Fig. 1, a simplified 5G handover procedure is given. Basically, there are general three stages of this. 
\subsubsection{Measurement and monitoring stage}
The TU is receiving the data packet from the Serving gNB via Unix to Unix (Uu) interface. The Serving gNB sends the Measurement Configuration (Config.) information regarding a re-connection message to the TU. The TU performs and processes the collected Received Signal Strengths (RSSs) measurements, and the Measurement Report (Rep.) is sent by TU to the Serving gNB. And the TU reports the measurements at specified time intervals (10 ms).

\begin{figure}[htbp]
	\centering
	\includegraphics[width=.9\linewidth]{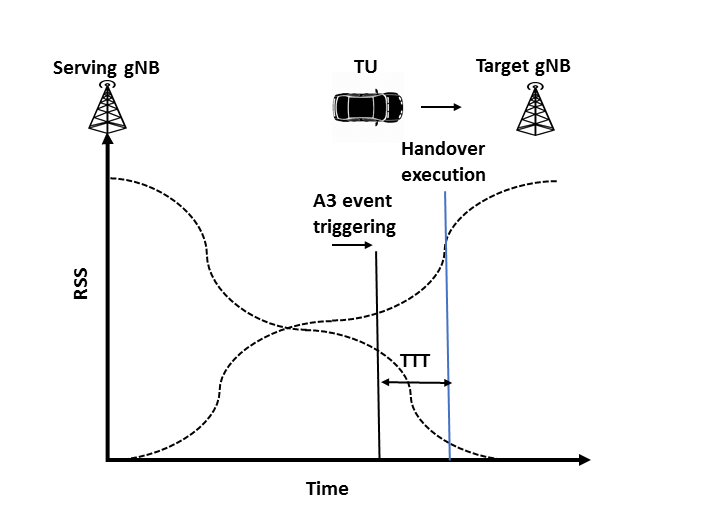}
	\caption{A3 event triggering}
	\label{fig}
\end{figure} 

\subsubsection{Handover Decision Making}
For making the handover decision, the A3 event [9] must be met, which is the entry condition to assess if the RSS of the Target gNB is better than the RSS of the current Serving gNB, in addition to a hysteresis margin (A3 margin: 3dB) as shown in Fig. 2. The A3 event has to be kept during a time period which is the TTT timer. Because handover reliability and handover frequency deeply rely on the TTT timer,  TTT is an important factor to be detected in the 5G NR handover. If all the above situations are met, then the Serving gNB makes the handover decision and sends a handover Request (Req.) to the Target gNB via Xn interface as X2/Xn as shown in Fig. 1. 
\subsubsection{Handover Execution Process}
The handover execution performs real handover steps. The wireless link between TU and Serving gNB is broken, and a new wireless link between TU and Target gNB is created. If the TU connects with the Target gNB successfully, the handover process is over. More real handover processing details can be found in [6]. 
\subsection{Algorithm for 5G handover-triggering}
In this part, the functions for the 5G handover triggering algorithm are provided. 

Step 1: The geometry (in dB) for TU at each tic is calculated which represents the signal quality, it measures the strength of the wanted signal compared to the unwanted interference and noise:

\begin{equation}
geo_i=10log_{10}  \left ({\frac{P_i}{\sum_{\mkern-5mu k \neq i} P_K +N_0 }} \right)   \label{eq}
\end{equation}

where $geo_i$ is the geometry of $TU_i$ at one tic w.r.t current serving\_gNB $i$, $P_i$ is the power received from the current serving\_gNB $i$. $K$ represents other gNBs except for gNB $i$. $P_K$ is the receiving power from other gNBs. The received power is calculated based on transmit power, pathloss (function of distance, frequency, or antenna heights), shadowing, fast-fading, and antenna gain as shown in Table II. In addition, only those cells covering the $TU_i$ are considered. The number of other cells is related to the density of gNBs and the communication ranges of gNBs which will be discussed in the scenario setting section. 

Step 2:
For $TU_i$, the $best\_geo$ at each tic, i.e., the biggest value of all geometries is calculated. 

\begin{equation}
best\_geo =  MAX \left ( geo_{i}, geo_{i+1}, geo_{i+2},... \right)   \label{eq}
\end{equation}

Then the target\_gNB position is obtained which has $best\_geo$. As the user starts moving, values of the TU's geometry start changing accordingly. The possibility of a user handover needs to be checked accordingly.

\begin{algorithm}
    \caption{5G handover triggering logical algorithm}\label{euclid}
    \hspace*{\algorithmicindent} \textbf{Input}: serving\_gnb, serving\_geo, best\_geo, target\_gnb, sinr\_min, avg\_geo, best\_cio, current\_cio, ho\_hys, TTT, ho\_timer, ho\_trigger, ho\_exec\_time    \\
    \hspace*{\algorithmicindent} \textbf{Output}: ho\_times
    \begin{algorithmic}[1]
    \If {$serving\_{gnb} \neq target\_{gnb}$} 
        \If{$best\_geo > sinr\_min  \&  
            best\_geo - avg\_geo + best\_cio - current\_cio) > ho\_hys$}
         \State  $ho\_trigger\gets  1$
         \State  $ho\_timer  \gets  ho\_timer +1$
            \If{$ho\_timer == TTT$}
             \State $serving\_gnb \gets  target\_gnb$
             \State $ho\_exec\_time\gets  25 $
             \State $ho\_times \gets ho\_times + 1 $
             \State $ho\_trigger \gets  0$
             \State $ho\_timer   \gets  0$
            \EndIf
         \EndIf
    \EndIf
    \end{algorithmic}
\end{algorithm}

The 5G handover logical algorithm is given in Algorithm. I in detail. And all used parameters regarding inputs and outputs are described in Table II. 
As shown in Algorithm. I, we want to get the total successful handover times in each simulation. The handover triggering logical algorithm comes after the following steps. 
1) it's to make sure the $target\_gnb$ is not as same as the current $serving\_gnb$ which means they locate in different positions; 2) it's to check the $best\_geo$ calculated before that should be greater than the pre-defined Signal-to-Interference-plus-Noise Ratio (SINR) threshold $sinr\_min = -7 dB$ and also to check the $best\_geo$ minuses $avg\_geo$, and $current\_cio$, pluses $best\_cio$ that should be greater than the handover hysteresis $ho\_hys = 3 \, dB$. $best\_cio$, and $current\_cio$ are 0 here, but this value will be affected when the load balancing algorithm is used. $avg\_geo$ is the average geometry of the UE w.r.t current gNB. It can be calculated using previous $X$ values and taking the average. In this simulation, a counter is used to hold 10 previous geometry values and $avg\_geo$ is calculated at each tic in the simulator. Specifically, it can be calculated only when the previous $X = 10$ values are available; 3) and 4) if all mentioned conditions are satisfied, then the $ho\_trigger$ flag is set to one and the $ho\_timer$ counter is added one; 5) it's to check whether the $ho\_timer$ fulfills the TTT value or not which is the key point that will be evaluated in this work because the TTT values affect a lot on the handover rate; 6) it's the execution step, $serving\_gnb$ equals to $target\_gnb$; $ho\_exec\_time$ equals to 25 tics; the output counter $ho\_times$ adds 1, and the $ho\_trigger$ and $ho\_timer$ reset to 0. In addition, if one of these mentioned conditions is not satisfied, the algorithm will rerun. The above description is the basic logic for the handover triggering in each simulation. 
\section{Simulation scenarios}
To check the 5G handover on variable TTTs and different UDNs, we built a downlink system-level simulator in Python. 
\subsection{gNB deployment}
As shown in Fig. 3, there are two simulation scenarios, where the gNB deployment in UDN is following the Poisson Point Process (PPP) [17]. Because only intra-handover is considered in this work, we assume all gNBs are the same with the same transmitting power and other same characteristics. In Fig. 3, an example of $den\_{gNB}$ of 20 is given which means there are 20 gNBs following PPP distribution in a 1000 m*1000 m urban area. In this paper, $den\_{gNB}$ of 10, 20, 30, 40, and 50 are used to analyze the effect of UDNs on 5G handover.
\subsection{User mobility model}
We set two different TUs running routes from different directions, starting and ending points. For example, in Fig. 3(a), the starting point and ending point are [1000,0] and [0,1000] separately with a speed of 50 km/h, and the direction angle ($\theta$) is 135 degrees.
Another simulation route detail can be found in Table I. Moreover, all other detailed simulation parameters are provided in Table II including physical layer and system layer information. 



\begin{table}[htbp]
\caption{User route models}
\begin{center}
\begin{tabular}{|p{0.9cm}|p{2cm}|p{2cm}|p{2.5cm}|}
 \hline
    Case A & starting point:[1000,0] & ending point:[0,1000] & direction:   $\theta = 135$\\
 \hline
    Case B & starting point:[1000,500] & ending point:[0,500] & direction:  $\theta = 180$\\
 \hline
\end{tabular}
\end{center}
\end{table}

\begin{figure}
    \centering
    \subfigure[Case A]{\includegraphics[width=0.24\textwidth]{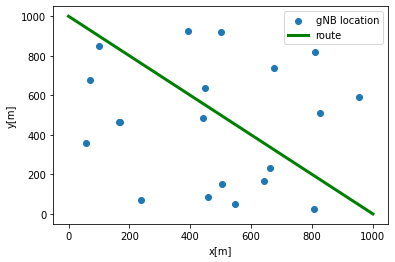}} 
    \subfigure[Case B]{\includegraphics[width=0.24\textwidth]{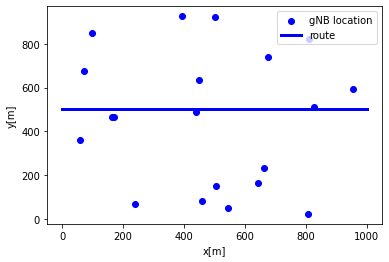}}
    \caption{Simulation scenarios with den\_gNBs(20) for different directions}
    \label{fig}
\end{figure}


\begin{table}[htbp]
\caption{Simulation parameters}
\begin{center}
\begin{tabular}{ |p{3cm}|p{5cm}|  }
 \hline
    \textbf{Parameters} &  \textbf{Description and Value} \\
 \hline
    Scenario & city in 1000 m*1000 m area \\
 \hline
    TU velocity& 13.89 m/s (50 km/h) \\
 \hline
    den\_gNB & density of gNBs per square kilometer(10, 20, 30, 40, 50)\\
 \hline 
   $\theta_i$ & direction of $TU_i$ (in degree) \\
 \hline 
   TU running time & 70000 ms \\
 \hline
    gNB height & 15 m\\
 \hline
    gNB coverage& 300 m\\
 \hline
    Carrier frequency& 6 GHz\\
 \hline
    bw & Bandwidths (10 MHz: 50 PRBs)\\
 \hline
    Transmit power & 30 dB \\
 \hline
    gNB antenna gain & 15 dBi\\
 \hline
    Receiver antenna gain & 0\\
 \hline
    Noise power &  -174 dBm/Hz + $10*\log_{10}$(bw) + 7\\
 \hline
    Pathloss model & pathloss = 128.1 + $37.6*\log_{10}$(Distance) \\
 \hline
    1 tic & 10 ms \\
 \hline
    serving\_gnb  & current serving gNB location \\
  \hline
    target\_gnb  & target gNB location \\
 \hline
    serving\_geo & the geometry of TU from current serving gNB \\
 \hline
    best\_geo &  the geometry of TU from the best connection gNB \\
 \hline
    sinr\_min & the minimum SINR value to keep TU connected to gNB (-7 dB)\\
 \hline 
    avg\_geo & average geometry of the TU w.r.t current gNB \\
 \hline
    ho\_avg\_geo\ & the average geometry value of TU for each successful handover (dB)\\
 \hline
    best\_cio & cell individual offset (0 dB) \\
 \hline
     current\_cio & cell individual offset (0 dB) \\
 \hline
    ho\_hys & handover hysteresis threshold (3 dB) \\
 \hline
    ho\_timer & handover counter \\
 \hline
    ho\_trigger & handover trigger flag (0 or 1)\\
 \hline
    ho\_exec\_time & handover execution time (25 tics)\\ 
 \hline 
    ho\_times & handover counter for summing the total times of handover \\
 \hline
\end{tabular}
\end{center}
\end{table}

\section{simulation results analysis}
The simulation results in this paper are analyzed to check the effect of UDNs and TTTs on the 5G NR handover times and performance. In general, we have 100 iterations for each simulation and get the average value for each time. 
\subsection{KPIs for measuring the handover effect}
We have two Key Performance Indicators (KPIs) to show the performance of the 5G UDN handover.
\subsubsection{Average handover times}
One is the average handover times per TU after each run which indicates the number of handover events that have been successfully executed. We call this KPI as handover rate. When the value of the average handover rate is less than 1, we assume it's a handover failure. 
\subsubsection{Average geometry value}
Another KPI we use it's the average geometry of a TU to represent the handover performance after each run which means the average geometry value of a TU after every successful handover execution

\begin{figure*}[htbp]
	\centering
	\subfigure[Case A]{
		\includegraphics[width=0.35\textwidth]{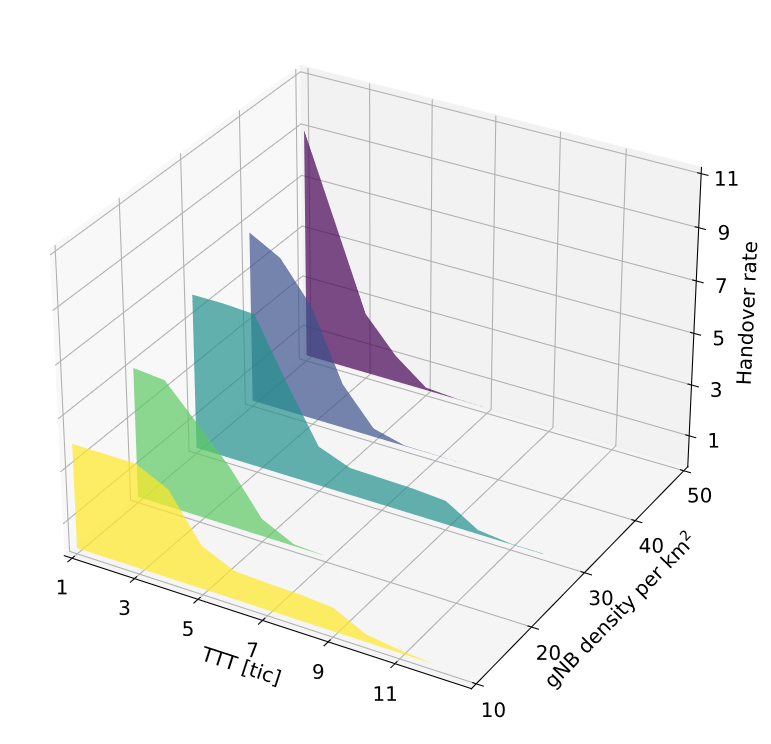}}
	\label{fig.sub.1}
	\subfigure[Case B]{
		\includegraphics[width=0.35\textwidth]{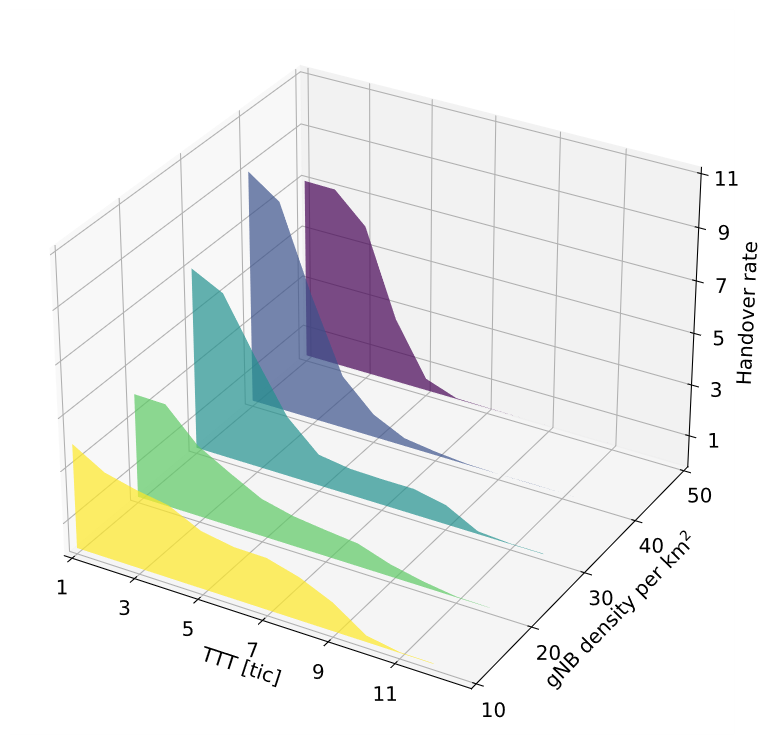}}
	\label{fig.sub.2}
    \caption{Simulation results for the four different scenarios}
    \label{fig}
\end{figure*}


\begin{equation}
\begin{aligned}
   ho\_avg\_geo = & MEAN (best\_geo_{ij},best\_geo_{ij+1},...,\\
   & best\_geo_{ij+k}) \label{eq}
\end{aligned}
\end{equation}
where $best\_geo_{ij}$ is the best geometry of $TU_i$ connected to $gNB_j$. There are $k$ times handover at run time, $k$ $best\_geo$ values are collected and the average geometry of $TU_i$ is achieved. 

\subsection{Effect of variable TTT values and density of UDN with fixed velocity on handover}
To get various simulation results, the range of TTTs is selected from 1 to 12 tics, the densities of gNBs are 10, 20, 30, 40, and 50, and velocity is fixed at 50 km/h. In Fig. 3, the 3-Dimension figures are correspondingly generated for the two assumed simulation scenarios.
From the simulation results, we can find almost similar simulation results for case A and case B, which means starting location or the direction of TU for different simulation scenarios doesn't affect the final results. 

In Fig. 4(a) case A, it's easy to find the overall handover rate decreases with the increasing TTT values when the density of gNB is 10. The rate of handover decreased from 4 to 0.01 with TTT increasing from 1 tic to 12 tics, which means if we want to have less handover rate,  the TTT value should be increased. But if the TTT value is largely increased such as $TTT=12$, the handover rate is below 1 as we defined there is no successful handover no matter how big the density of gNB is. Because the TU will experience a severe degradation of geometry during the TTT period. This indicates, that a large TTT value will lead to handover failure, it's important to know the effect of the TTT value on the handover rate and to select a proper TTT value.

In Fig. 4(b) case B, when the TTT value is 1, with the increasing density of gNB from 10 to 50, the handover rate is growing from 4 to 8.97 which means the increasing density of UDN will largely raise the handover rate when the TTT value is not large. Ultra-dense deployment of gNB leads to redundant handovers. The density of UDN doesn't affect the handover rate anymore if TTT is larger than 8. The larger the TTT for the handover procedure, the less effect of density of UDN has on the handover times.

Overall, finding a proper TTT value for the use case is vital for the overall handover performance.

\begin{table}[htbp]
\caption{Average geometry of TU for Case A}
\begin{center}
\begin{tabular}{ |p{3cm}|p{0.5cm}|p{0.5cm}|p{0.5cm}|p{0.5cm}|p{0.5cm}|  }
\hline
\backslashbox[35mm]{\textbf{TTT\_ms}}{\textbf{ho\_avg\_geo}}{\textbf{density}}
&\makebox{10}&\makebox{20}&\makebox{30}
&\makebox{40}&\makebox{50} \\
 \hline
    1 & 16.22 & 6.54  & 8.06  & 4.28   & 5.75\\
 \hline
    2 & 16.97 & 7.44  & 9.17  & 5.2    & 7.6\\
 \hline 
    3 & 22.54 & 10.58 & 13.33 & 8.23   & 11.49\\
 \hline 
    4 & 33.19 & 12.97 & nan   & nan    & 18.81\\
 \hline
    5 & 38.44 & nan   & nan   & nan    & nan\\
 \hline
    6 & 39.76 & nan   & nan   & nan    & nan\\
 \hline
    7 & 40.9  & nan   & nan   & nan    & nan\\
 \hline
    8 & 40.35 & nan   & nan   & nan    & nan\\
 \hline 
    9 & nan   & nan   & nan   & nan    & nan \\
 \hline
    10 & nan  & nan   & nan   & nan    & nan\\
 \hline
    11 & nan  & nan   & nan   & nan    & nan\\
 \hline
    12 & nan  & nan   & nan   & nan    & nan\\
 \hline
\end{tabular}
\end{center}
\end{table}

\subsection{Performance on handover}
For more in-depth research and to get an optimum TTT value, we get the tables of average geometry values for two scenarios to show the handover performance which can clearly tell us the influences of TTT and the density of UDN on the performance of handover. In these two tables, TTT, density, and the ho\_avg\_geo are given.

In Table III for case A, if the density of UDN is 10, it's easy to tell when the value of TTT is 8 with 1 handover rate from Fig. 4(a), the average handover performance is the best which is 40.35 dB. But if the TTT value is continually increased, there are non-values of handover performance, because too large a TTT value will lead the TU to lose the connection with the network, which means no successful handover has been made under this situation. With increasing the density of UDN, less useful TTT values can be chosen for efficient handover performance. When the density is 20, the best TTT is 4. But for the density of 30 and 40, the best TTT value is 3. In this case, when $density > 20$, TTT selection can only be less than or equal to 4, which means the larger density of UDN, the more strict picking of the TTT values.

From Table IV case B, simulation results are a slight difference compared to the same simulation parameters in case A. When the density is 10, the best TTT value is 7 and the average geometry is 30.36 dB. if the density is raised to 20, the best TTT value is 6 and the average geometry is decreased to 29.46 dB. From both case A and case B for $TTT=1$, the performance of the TU at a high density of UDN dropped a lot compared to the performance of the TU at a low density from 16.22 dB to 5.75 dB in case A and from 16.93 dB to 4.04 dB. Due to the increased density of gNBs, the TU will not only face many times handover but also more intense signal interference from nearby gNBs. That's why the performance of the TU decreases with the increasing density of gNBs. The two tables are able to be utilized as mapping tables which it's easy to get the best TTT values for different use cases.





\begin{table}[htbp]
\caption{Average geometry of TU for Case B}
\begin{center}
\begin{tabular}{ |p{3cm}|p{0.5cm}|p{0.5cm}|p{0.5cm}|p{0.5cm}|p{0.5cm}|  }
\hline
\backslashbox[35mm]{\textbf{TTT\_ms}}{\textbf{ho\_avg\_geo}}{\textbf{density}}
&\makebox{10}&\makebox{20}&\makebox{30}
&\makebox{40}&\makebox{50} \\
 \hline
    1 & 16.93 & 4.05 & 4.11  & 5.2    & 4.04\\
 \hline
    2 & 17.93 & 6.50 & 5.86  & 6.38   & 5.45\\
 \hline 
    3 & 29.73 & 11.78 & 10.36 & 10.11  & 9.82\\
 \hline 
    4 & 32.28 & 19.22  & nan   & nan    & nan\\
 \hline
    5 & 29.93 & 26.88 & nan   & nan    & nan\\
 \hline
    6 & 30.14 & 29.46 & nan   & nan    & nan\\
 \hline
    7 & 30.36 & nan & nan   & nan    & nan\\
 \hline
    8 & nan   & nan & nan   & nan    & nan\\
 \hline 
    9 & nan   & nan   & nan   & nan    & nan \\
 \hline
    10 & nan  & nan   & nan   & nan    & nan\\
 \hline
    11 & nan  & nan   & nan   & nan    & nan\\
 \hline
    12 & nan  & nan   & nan   & nan    & nan\\
 \hline
\end{tabular}
\end{center}
\end{table}

\subsection{Effect of variable TTT values and velocity with a fixed density of UDN on handover}

\begin{figure*}[htbp]
	\centering
	\subfigure[handover rate for different velocities and TTT values ]{
		\includegraphics[width=0.45\textwidth]{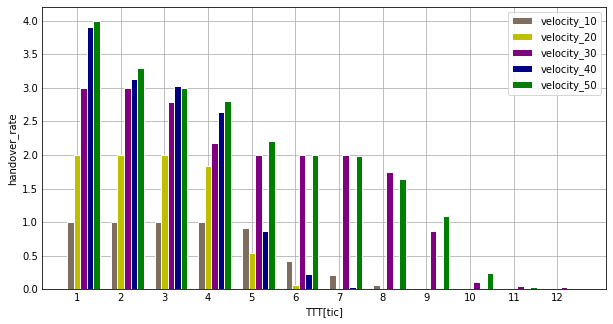}}
	\label{fig.sub.1}
	\subfigure[Average geometry for different velocities and TTT values]{
		\includegraphics[width=0.45\textwidth]{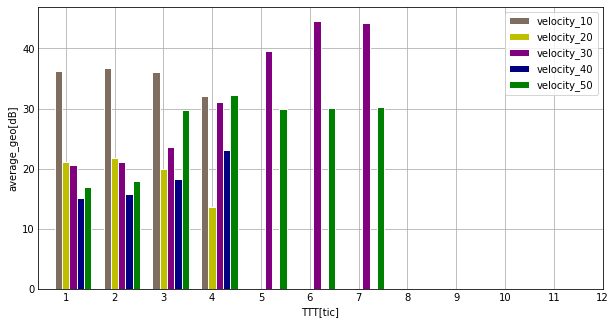}}
	\label{fig.sub.2}
    \caption{Simulation results for different velocities and TTT values for Case B}
    \label{fig}
\end{figure*}
In this part, the simulation results for evaluating the effect of variable TTT values and velocities with a fixed density of UND on handover performance are given in Fig. 5.
The testing use case is case B from Table I. Velocities are 10,20,30,40, and 50 km/h. The fixed density of UDN is 10. First of all, in Fig. 5(a), when the TTT value is 1, the handover rate is increasing from 1 at 10 km/h to 4 at 50 km/h, this is because fast-moving TU will face more chances to have better gNBs' connections compared to the slow-moving car in the same running time. But multiple handover times in a short time don't mean the average performance of handover will be good enough which can be found in Fig. 5(b), the average geometry drops from 36.29 dB to 16.93 dB correspondingly.

In Fig. 5(b), when a TU moves at 50 km/h, with increasing TTT values, the average geometry rises from 16.93 dB at $TTT = 1$ to 30.3 dB at $TTT =7$. When TTT is between 4 and 8, the optimal velocity of 30 km/h provides the best handover performance. 
 
\section{Conclusion}
This paper analyzes the effect of TTT values, different densities of UDNs, and different velocities of TUs on the 5G handover process. A simulator is developed to evaluate the handover times and handover performance for different simulation scenarios in Python. Simulation results show that the TTT value has a large effect on the handover rate. Too large TTT values will decrease the handover times but make TUs lose network connection. However, too small TTT will lead to too frequent and unnecessary handovers. Moreover, the UDN is an important solution to raising network capacity and data traffic in 5G. According to the simulation results, the density of UDN also has a large influence on the handover performance. There must be a compromise between the TTT and the density of UDN. The handover performance mapping tables for evaluating the effect of TTT values and the density of UDN on the performance of TU on 5G handover are given. According to the tables, it's easy to see how to select the proper TTT values and densities to optimize the handover performance for different use cases. Additionally, the handover performance of different TTT values and velocities are specified, by the combination of TTT values and TU's velocity, the optimized handover performance is gained as well. 

For our future work, some more coming research will be done to reduce the handover rate and select proper TTT values, the density of UDN, or proper velocities of TU by applying machine learning algorithms on 5G wireless and maximizing the 5G communication performance.

\end{document}